\documentclass[12pt]{article}
\usepackage{epsf}
\usepackage{graphicx}
\usepackage{amssymb}
\usepackage{psfrag}

\def\hybrid{\topmargin 0pt      \oddsidemargin 0pt
	\headheight 0pt \headsep 0pt
	\textheight 9in         
 	\textwidth 6.25in       
	\marginparwidth .875in
	\parskip 5pt plus 1pt   \jot = 1.5ex}

\catcode`\@=11

\def\numberbysection{\@addtoreset{equation}{section}
\def\theequation{\thesection.\arabic{equation}}}

\numberbysection
\hybrid

\begin{document}

\begin{titlepage}
\begin{center}
December~2008 \hfill . \\[.5in]
{\large\bf A lattice model for the second $\mathbb{Z}_{3}$ parafermionic field theory} 
\\[.5in] 
{\bf Benoit Estienne}\\[.2in]
 {\it LPTHE, CNRS, UPMC Univ Paris 06 }\\

 {\it Bo\^{\i}te 126, 4 place Jussieu, F-75252 Paris Cedex 05}\\
estienne@lpthe.jussieu.fr \\

\end{center}

\underline{Abstract.}

The second $\mathbb{Z}_{3}$ parafermionic conformal theories are associated with the coset construction $\frac{SU(2)_{k}\times SU(2)_{4}}{SU(2)_{k+4}} $.  Solid-on-solid integrable lattice models obtained by fusion of the model based on level-1 representation of the affine algebra $B_1^{(1)}$ have a critical point described by these conformal theories. Explicit values for the Boltzmann weights are derived for these models, and it is shown that the Boltzmann weights can be made positive for a particular value of the spectral parameter, opening a way to eventual numerical simulations of these conformal field theories. Away from criticality, these lattice models describe an integrable, massive perturbation of the parafermionic conformal theory by the relevant field $\Psi_{-2/3}^{\dagger}D_{1,3} $.

\end{titlepage}

\newpage

Over the past twenty years Conformal Field Theories (CFT) have proven to be a very powerful tool to describe two dimensional statistical systems at criticality. Recently a new way to study critical systems has been developed : Schramm Loewner Evolution (SLE) \cite{ref1,ref2}. In this approach to conformally invariant systems, clusters are described through the conformal properties of their domain walls. While the connection between minimal models ($c<1$) and SLE is relatively well understood, it is still an open problem for CFTs possessing extra symmetries. Statistical systems with a discrete symmetry $\mathbb{Z}_N$ are believed to be described by parafermionic theories. The first of such theories, constructed by Fateev and Zamolodchikov \cite{ref3}, have been studied in the context of SLE \cite{ref4,refraoul}, and using particular lattice realizations of the first parafermions, numerical studies of fractal interfaces have been performed \cite{refgamsa,refnumerics,refnumerics2}.

Conformal theories associated with the coset construction $\frac{SO(N)_{k}\times SO(N)_{2}}{SO(N)_{k+2}} $  ($N\geq 5$) are referred to as (second) parafermionic CFTs $\mathbb{Z}_{N}^{(2)}(k)$ \cite{ref6,ref7,ref8,ref9}. 
Having a lattice version of these parafermions is very appealing for several reasons. It would allow a better understanding of this specific $\mathbb{Z}_N$ symmetry, how it can be realized, and how it's broken when the phase transition occurs. And it gives access to the related SLEs, at least numerically through the study of interfaces at criticality. 

Using the coset description of the parafermions, techniques to build integrable lattice models are available : the fusion procedure \cite{ref10,ref11} applied to particular existing lattice models \cite{ref12}. In general the models we end up with have fluctuations variables both on sites and edges of a square lattice, and local interactions are around faces. When $N=3$ this general picture is somewhat simpler, since there is no edge variable in that case. The  corresponding restricted solid-on-solid (RSOS) models have been studied in the context of solvable models obtained by fusion of the 8 vertex model \cite{ref13,ref14,ref15}. In particular their critical behavior has been identified by evaluation of the local state probability (LSP), and it is known that they are described by the $N=3$ second parafermions.

The purpose of this article is to study an explicit realization of these $\mathbb{Z}_{3}^{(2)}$ second parafermions \cite{ref5}. We adopt a somewhat different point of view, having in mind a generalization toward the case $N \geq 3$, but up to some conventions the lattice model we get is strictly equivalent to those described above. The focus of the paper is on practical aspects of this model, in particular it is shown that all Boltzmann weights can be made positive for a particular value of the spectral parameter, opening a way to numerical simulations.

\section{Description of the model}

Starting from the coset description of the parafermions : $\mathbb{Z}_{N}^{(2)}(k) = \frac{SO(N)_{k}\times SO(N)_{2}}{SO(N)_{k+2}} $, the standard way to build an integrable lattice model is by fusion of the model realizing $\frac{SO(N)_{k}\times SO(N)_{1}}{SO(N)_{k+1}}$. These JMO models \cite{ref12} are related to the vector representation of $SO(N)$. 

For $N=3$ it is known that this is equivalent to the model corresponding to the symmetric tensor representation of degree 2 of $SU(2)$, which in turns is obtained by fusion of the model based on the vector representation of $SU(2)$. This last model is nothing but the celebrated ABF model \cite{ref16}.

There are two equivalent ways to obtain a solvable lattice models realizing $\mathbb{Z}_{3}^{(2)}$ : by a 4-fusion of the vector $SU(2)$ model, or by a 2-fusion of the vector $SO(3)$ model. In terms of CFT, this is related to the following coset equivalence :

\begin{equation}
\mathbb{Z}_{3}^{(2)}(k)=\frac{SU(2)_{k}\times SU(2)_{4}}{SU(2)_{k+4}}  =  \frac{SO(3)_{k/2}\times SO(3)_{2}}{SO(3)_{k/2+2}}
\end{equation}

As a coset based on $SU(2)$, integrable lattice models obtained by 4 fusions of the ABF model are readily available \cite{ref13,ref15}. 

Equivalently, this can be seen as a 2 fusion of the model based on vector representation of $SO(3)$, and this is the point of view adopted here, since it will be closer to the general case ($N \geq 3$). This will lead to some different conventions from the usual $SU(2)$ lattice models, in particular we will have to allow half integer values for the heights and for the ``level'' $k/2$, but it should be stressed that both models are exactly the same.

\begin{figure}
 \centering
 \psfrag{1}[0][0][1][0]{$1$}
 \psfrag{2}[0][0][1][0]{$2$}
 \psfrag{3}[0][0][1][0]{$3$}
 \psfrag{4}[0][0][1][0]{$4$}
 \psfrag{5}[0][0][1][0]{$5$}
 \psfrag{6}[0][0][1][0]{$6$}
 \psfrag{7}[0][0][1][0]{$7$}
 \psfrag{8}[0][0][1][0]{$8$}
 \includegraphics[scale=0.7]{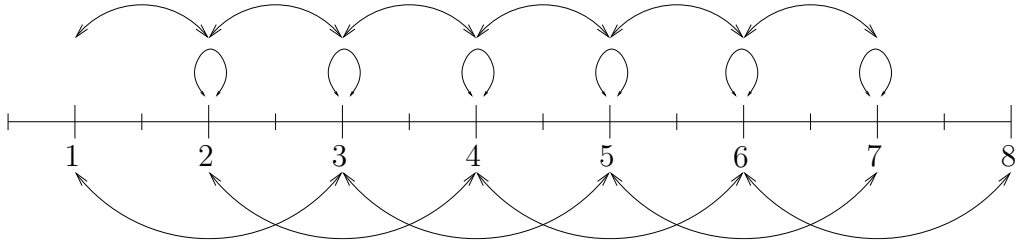}
 \caption{Nearest neighbour constraints for $L=17/2$. Only the integer heights are shown. Arrows indicate admissible pairs.\label{admissibility}}
\end{figure}

Consider a two dimensional square lattice, with heights $a,b,c,d, \  etc$ at each site taking the values $1/2,1,...,L-1/2$. These local states are subject to nearest neighbour constraints : if $a$ and $b$ live on neighbouring sites, the pair $(a,b)$ must be admissible, i.e. $a-b=\pm 2,\pm 1,0$ and $2<a+b<2L-2$ (see Fig. \ref{admissibility} for an exemple). For this model to have any internal freedom, the parameter $L$ must be large enough : $L\geq 7/2$. In fact $L$ will be related to $k$ by $L=k/2+3$, so that the tricritical three state potts model correspond precisely to $L=7/2$. It turns out that due to the nearest neighbour constraints, for a given configuration the heights will be either all integer, or all half integer. We end up with two models for a given value of $L$, depending on whether we consider integer or half integer values for the heights.

\begin{figure}
 \centering
 \psfrag{a}[0][0][1][0]{$a$}
 \psfrag{b}[0][0][1][0]{$b$}
 \psfrag{c}[0][0][1][0]{$c$}
 \psfrag{d}[0][0][1][0]{$d$}
 \psfrag{A}[0][0][1][0]{$A$}
 \psfrag{B}[0][0][1][0]{$B$}
 \includegraphics[scale=0.4]{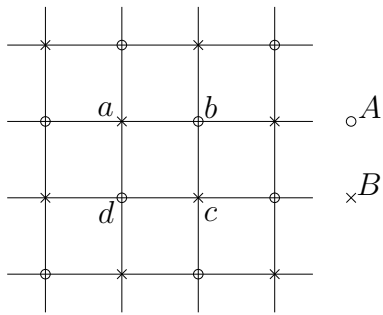}
 \caption{Four sites surrounding a face. The sublattices A and B are shown. \label{face}}
\end{figure}

The restricted solid-on-solid model is a face model, in the sense that the heights variables interact around faces of the square lattice. The statistical weight assigned to an elementary face of the lattice is zero unless all four pairs of adjacent heights on the edges are admissible. We will denote these weights as  $W \left( \begin{array}{ccc} 
a & b  \\
d & c 
\end{array}  \right)$ where $a$, $b$, $c$, $d$ are the four surrounding sites, ordered clockwise from the upper left, as
in Fig. \ref{face}.

Such a model is insensitive to the following gauge transformation of the weights, since the additionnal edge interactions will cancel out on two neighbouring faces : 

 \begin{equation}
W \left( \begin{array}{cc}
a & b  \\
d & c 
\end{array}  \right) \rightarrow \frac{f(a,b)g(a,d)}{f(d,c)g(b,c)} W \left( \begin{array}{cc}
a & b  \\
d & c 
\end{array}  \right) \quad \quad \forall f,g \neq 0  \end{equation}

To be more precise, the partition function on a torus (i.e. with periodic boundary conditions) is unchanged. If fixed boundary conditions are applied on the boundary, the partition function is simply multiplied by an irrelevant factor. This is typically the case when calculating the local state probability. If one is interested in more general boundary conditions, such as open boundary conditions, then these gauge transformations have a non trivial effect, localized at the boundary of the system. In the bulk, the model is insensitive to these gauge transformations.

Such transformations will be required to ensure the positivity of the Boltzmann weights (cf section \ref{signs}).

It should be noted that these gauge transformations preserve integrability : if the weights obey the star-triangle equation, they will still satisfy it after transformation.

Explicit values for the Boltzmann weights are obtained by fusion procedure of the original $SO(3)$ model. These are given in the appendix \ref{explicit_weights}.

The weights $W \left( \begin{array}{cc}
a & b  \\
d & c 
\end{array} \vline \, u,p  \right)$ depend on two parameters $u,p$. In the context of exactly solvable models, they are usually called spectral parameter, and nome (respectively). Whereas the spectral parameter $u $ plays a crucial role in the integrability of the model, it can be fixed once and for all for numerical simulations. The nome $0 \leq p \leq 1 $ plays the role of temperature. The system is critical for $p=0$, and at zero temperature for $p=1$. We will consider here only the so-called regime III, where $-1/2<u<0$ and $0 \leq p \leq 1$.

\begin{figure}
 \centering
 \psfrag{a}[0][0][1][0]{$a$}
 \psfrag{b}[0][0][1][0]{$b$}
 \psfrag{c}[0][0][1][0]{$c$}
 \psfrag{d}[0][0][1][0]{$d$}
 \psfrag{e}[0][0][1][0]{$e$}
 \psfrag{f}[0][0][1][0]{$f$}
 \psfrag{g}[0][0][1][0]{$g$}
 \psfrag{ap}[0][0][1][0]{$a$}
 \psfrag{bp}[0][0][1][0]{$b$}
 \psfrag{cp}[0][0][1][0]{$c$}
 \psfrag{dp}[0][0][1][0]{$d$}
 \psfrag{ep}[0][0][1][0]{$e$}
 \psfrag{fp}[0][0][1][0]{$f$}
 \psfrag{gp}[0][0][1][0]{$g$}
 \psfrag{u}[0][0][1][0]{$u$}
 \psfrag{up}[0][0][1][0]{$u$}
 \psfrag{v}[0][0][1][0]{$v$}
 \psfrag{vp}[0][0][1][0]{$v$}
 \psfrag{u+v}[0][0][1][0]{$u+v$}
 \psfrag{up+vp}[0][0][1][0]{$u+v$}
 \psfrag{=}[0][0][1][0]{$=$}
 \psfrag{sigma}[0][0][2][0]{$\sum_g$}
 \psfrag{sigmap}[0][0][2][0]{$\sum_g$}
 \includegraphics[scale=0.4]{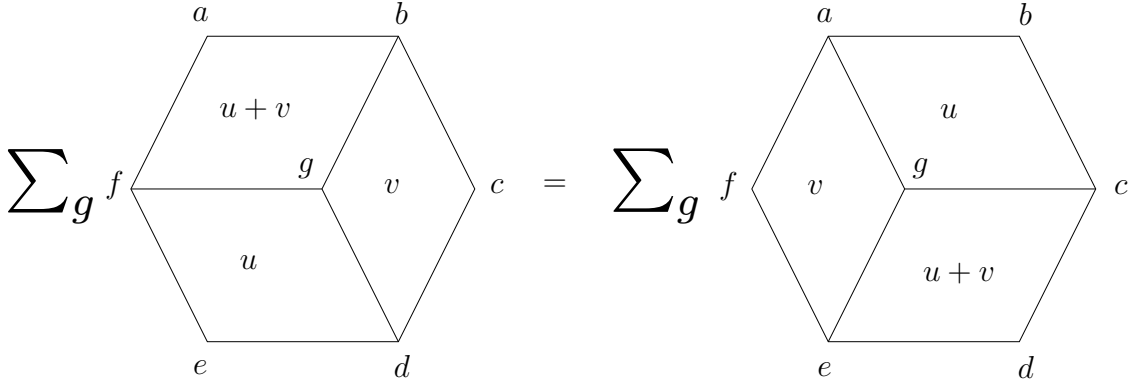}
 \caption{The star-triangle equation. The nome $p$ does not a play a role in the STR and has been omitted : it is the same for all faces. \label{YB}}

\end{figure}

By construction, the fusion procedure preserves many interesting properties from the original model, the most important of them being integrabililty : the weights of the fused model obey the \emph{star-triangle relation} (STR) Fig. \ref{YB}. The STR ensures that, in the case of periodic boundary conditions, row-to-row transfer matrices for different value of the spectral parameter $u$ commute, the nome $p$ being fixed.

\begin{eqnarray}
 & & \sum_g W \left( \begin{array}{cc}
f & g  \\
e & d 
\end{array} \vline \, u,p  \right) W \left( \begin{array}{ccc}
b & c  \\
g & d 
\end{array} \vline \, v,p  \right) W \left( \begin{array}{cc}
a & b  \\
f & g 
\end{array} \vline \, u+v,p  \right) \cr
& = & \sum_g W \left( \begin{array}{cc}
a & b  \\
g & c 
\end{array} \vline \, u,p  \right)W \left( \begin{array}{ccc}
a & g  \\
f & e 
\end{array} \vline \, v,p  \right)W \left( \begin{array}{cc}
g & c  \\
e & d 
\end{array} \vline \, u+v,p  \right) 
\end{eqnarray}

The Boltzmann weights obtained by fusion enjoy the following properties, herited from the original model : 

\begin{itemize}
 \item \emph{Reflection symmetry : }

 \begin{equation}W \left( \begin{array}{ccc}
a & b  \\
d & c 
\end{array} \vline \, u,p  \right) = W \left( \begin{array}{ccc}
a & d  \\
b & c 
\end{array} \vline \, u,p  \right) = W \left( \begin{array}{ccc}
c & b  \\
d & a 
\end{array} \vline \, u,p  \right) \end{equation}

\item \emph{Rotational symmetry :} 

\begin{equation}W \left( \begin{array}{ccc}
a & b  \\
d & c 
\end{array} \vline \, u,p  \right) = \left(\frac{G_b G_d}{G_a G_c}\right)^{1/2} W \left( \begin{array}{ccc}
d & a  \\
c & b 
\end{array} \vline \, \lambda-u,p  \right) \label{rotational}\end{equation}

\item \emph{$\mathbb{Z}_2$ symmetry :}

\begin{equation}W \left( \begin{array}{ccc}
a & b  \\
d & c 
\end{array} \vline \, u,p  \right) = W \left( \begin{array}{ccc}
L-a & L-b  \\
L-d & L-c 
\end{array} \vline \, u,p  \right) \end{equation}

\end{itemize}

where the following notations are used : $\lambda=-1/2$, $G_a=[a]$, and $[x]$ denotes an elliptic theta function (cf Appendix \ref{explicit_weights}).

From the rotational symmetry it appears that the spectral parameter $u$ is related to the spatial anisotropy of the interaction. The model is istropic for $u=\lambda/2=-1/4$, since the additionnal terms $\left(\frac{G_b G_d}{G_a G_c}\right)^{1/2}$ in equation (\ref{rotational})  can be absorbed in a gauge transformation (cf section  \ref{signs}).

\section{Some useful gauges transformations}
\label{signs}

From the values in the appendix, it appears that some Boltzmann weights are not positive at $u=-\lambda/2$. But in fact these signs can all be absorbed in the following gauge transformation : 

\begin{eqnarray}
 W \left( \begin{array}{cc}
a & b  \\
d & c 
\end{array}  \right) & \rightarrow & \frac{\epsilon(a,b)\epsilon(a,d)}{\epsilon(d,c)\epsilon(b,c)} W \left( \begin{array}{cc}
a & b  \\
d & c 
\end{array}  \right) 
\end{eqnarray}

where $ \epsilon(a,b) = \epsilon(b,a)$ is just a sign, given by (we consider here the case when the heights are integers) :

\begin{eqnarray}
\epsilon(a,a) & = & (-1)^{a} \cr
\epsilon(a,a+1) & = & (-1)^{a(a+1)/2} \cr
\epsilon(a,a+2) & = & 1 
\end{eqnarray}

This works with half integer just by shifting the heights by $1/2$, defining $\epsilon(a,b)=\epsilon(a-1/2,b-1/2)$ when $a,b$ are half integers.

The only effect of this gauge transformation is to change the sign of the following weights, and their symmetric obtained by reflexion and rotation  :

\begin{eqnarray}
 W \left( \begin{array}{cc}
a & a  \\
a \pm1 & a \pm2 
\end{array}  \right), &  W \left( \begin{array}{cc}
a & a  \\
a \pm1 & a \pm1 
\end{array}  \right), &  W \left( \begin{array}{cc}
a & a \pm1  \\
a  \pm2 & a \pm3 
\end{array}  \right)
\end{eqnarray}

In particular, this gauge preserve all the symmetries (reflexion, rotation, $\mathbb{Z}_2$) of the Boltzmann weights, and ensures their positivity for $u=-1/4$. The model is then both isotropic and positive for this specific value of the spectral parameter. Finally, to make the model explicitly isotropic, the following gauge transformation, with $f(a,b)=\left(G_a/G_b\right)^{1/4}$, $g(a,b)=1$, can be performed :

\begin{eqnarray}
 W \left( \begin{array}{cc}
a & b  \\
d & c 
\end{array}  \right) & \rightarrow & \tilde{W} \left( \begin{array}{cc}
a & b  \\
d & c 
\end{array}  \right) = \left(\frac{G_a G_c}{G_b G_d}\right)^{1/4} W \left( \begin{array}{cc}
a & b  \\
d & c 
\end{array}  \right) 
\end{eqnarray} 

This preserve $\mathbb{Z}_2$ and reflexion symmetries, and positivity, since $G_a = [a] > 0$ for all heights. And it makes the rotational symmetry more explicit :

\begin{eqnarray}
 \tilde{W} \left( \begin{array}{cc}
a & b  \\
d & c 
\end{array} \vline u=-1/4 \right) & = &  \tilde{W} \left( \begin{array}{cc}
b & c  \\
a & d 
\end{array} \vline u=-1/4 \right) 
\end{eqnarray} 

Once the spectral parameter is fixed at the isotropic value $u=-1/4$, the only remaining parameters of the model are the nome $p$, $0 \leq p \leq 1$, and the boundary on the heights $L$. If one is interested in the critical behavior of the model, the nome $p$ must be put to $0$, in which case the Boltzmann simplify considerably : all theta functions $[x]$ merely become  $\textrm{sin}(\pi x/L)$. To sum up, the model is critical and isotropic for $u=-1/4$ and $p=0$. The only parameter is then $L=k/2+3$, and correspond to the choice of the CFT describing the continuum limit of  this model :  $\mathbb{Z}_3^{(2)}(k)$, $k \geq 1$.

\section{Low temperature limit}

The low temperature limit correspond to $p\rightarrow 1$. The following reparametrization of the elliptic theta function is very useful to study this limit : in regime III, the weights can be reexpressed using the parameter $x$ rather than $p$, where :

\begin{eqnarray*}
p=e^{-\epsilon} & & x=e^{-4 \pi  / L \epsilon} \\
p\rightarrow 1 & & x\rightarrow 0
\end{eqnarray*}

To do so we use the following modular transformation :

\begin{eqnarray*}
[a] & = & \tau(x) x^{\frac{(a-L/2)^2}{2L}} E(x^a,x^L) \\
 E(z,x)& = & \prod_{n=1}^{\infty}\left(1-zx^{n-1}\right)\left(1-z^{-1}x^{n}\right)\left(1-x^{n}\right) \\
\end{eqnarray*}

where $\tau(x)$ is a certain function of $x$ here, irrelevant here because it cancels out when evaluating ratios of theta functions. The zero temperature limit is now $x \rightarrow 0$. This is very convenient to study the low temperature properties of the model, because the product expansion for $E(z,x)$ is rapidly convergent as $x\rightarrow 0$.

In particular the limit $x\rightarrow 0$, $w=x^u$ fixed, is very interesting since the weights become diagonal :

$$\lim_{x \rightarrow 0,w = cst} W \left( \begin{array}{cc}
a & b  \\
d & c 
\end{array} \vline \, u,p \right) = F(u) \delta_{b,d} w^{-H(a,b,c) + f_a+f_c-f_b-f_d}$$

where 
\begin{eqnarray*}
F(u)  & = & x^{\frac{2u^2+u}{L}} \\
f_a & = & \frac{|a-L/2|^2}{2L}  \\
H(a,b,c)& = & \frac{|a-c|}{2}
 \end{eqnarray*}

In this expression $F(u)$ is just a global multiplicative factor, and $w^{f_a+f_c-f_b-f_d}$ can be absorbed in a gauge transformation. What remains is :

\begin{equation}
\lim_{x \rightarrow 0,w \textrm{ fixed}} W \left( \begin{array}{cc}
a & b  \\
d & c 
\end{array} \vline \, u,p \right) \sim \delta_{b,d} w^{-H(a,b,c)}\end{equation}

In this limit, only configurations invariant under translations along the southwest to northeast diagonal contribute to the partition function. Thus what is left is an effective one-dimensional problem. Using the corner transfer matrix trick, one can use this one-dimensional limit to calculate exactly one point functions (cf appendix \ref{LSP}). Let us proceed to examine the ground states of this model.

Ground states are configurations that maximize the Boltzmann weights (in absolute value). In regime III, which correspond to the region $-1/2<u<0$, $0 \leq p \leq 1$, the ground states do not depend on the value of $u$ and $p$. The one-dimensional limit is very practical to extract information about them. One can readily see that all ground states must be invariant under translations along the southwest to northeast diagonal. Then to maximize the term $w^{-H(a,b,c)}$, we must take $(a,b,c)$ such as $H(a,b,c)$ is minimal : this is acheived for $a=c$. Ground states are thus invariant under translations along the southeast to northwest diagonal as well.

This is consistent with the fact that for the regime III, ground states must be isotropic, since one could have been interested in the limit where $x\rightarrow 0$, while $\tilde{w}=x^{\lambda-u}$ is fixed. In that limit the weights would behave as (up to some gauge transformation):

\begin{equation}\lim_{x \rightarrow 0,\tilde{w} \textrm{ fixed}} W \left( \begin{array}{cc}
a & b  \\
d & c 
\end{array} \vline \, u,p \right) \sim \delta_{a,c} \tilde{w}^{-H(d,a,b)}\end{equation}

\begin{figure}
 \centering
 \psfrag{b}[0][0][1][0]{$b$}
 \psfrag{c}[0][0][1][0]{$c$}
 \psfrag{A}[0][0][1][0]{$A$}
 \psfrag{B}[0][0][1][0]{$B$}
 \includegraphics[scale=0.7]{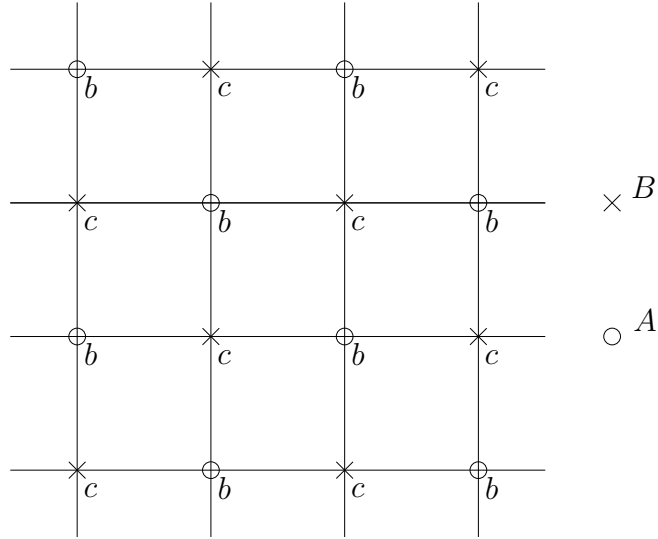}
 \caption{Ground state $(b,c)$. Heights living on the sublattice $A$ assume the value $b$, while those on the sublattice $B$ take the value $c=b+\eta$ \label{ground_state}}
\end{figure}

Every configuration invariant under translations along diagonals is a ground state. Decomposing the square lattice in sublattices $A$ and $B$ (see Fig. [\ref{ground_state}]), a ground state will be described by a pair of admissible heights $(b,b+\eta)$, such that heights assume the value $b$ on the sublattice $A$, and $c=b+\eta$ on $B$. Thus there are five types of ground states :
\begin{itemize}
 \item $(b,b)$, for $3/2 \leq b \leq L-3/2$ 
 \item $(b,b+1)$ and $(b+1,b)$ for $1 \leq b \leq L-2$
 \item $(b,b+2)$ and $(b+2,b)$ for $1/2 \leq b \leq L-5/2$
\end{itemize}

The boundaries on $b$ come from the admissibility rules for neighbouring sites. A compact way of writing these constraints is : $3/2 \leq \bar{b} \leq L-3/2$. where $\bar{b}$ is just the average height value in the ground state $(b,c)$ : $\bar{b}=\frac{b+c}{2}$.
It should be stressed that $(b,c)$ and $(c,b)$  are different ground states (for $c\neq b$).

It is interesting to note that there is no obvious realization of the $\mathbb{Z}_3$ symmetry in the structure of the ground states.

\section{Critical behavior and Local State Probability}

The model becomes critical as $p \rightarrow 0$. Exploiting the integrability of the system, it is possible to extract the critical exponents exactly. The inversion method \cite{refBaxter} gives a method to calculate the critical exponent $\alpha$. Unfortunately it turns out that the definition of $\alpha$ is ambiguous for this model, since the free energy per site is regular as $p$ goes to $0$, as was observed in \cite{ref13}. Anyhow, in view of the form of the local state probability in equation (\ref{LSP_expr}), there is only one natural definition for $\alpha$ consistent with the relation between critical exponents and conformal dimensions of the CFT : 

\begin{equation}
2-\alpha  = L  
\end{equation}

\begin{figure}
 \centering
 \psfrag{a}[0][0][1][0]{$a$}
 \psfrag{b_m}[0][0][1][0]{$b_m$}
 \psfrag{b_{m+1}}[0][0][1][0]{$b_{m+1}$}
 \includegraphics[scale=0.4]{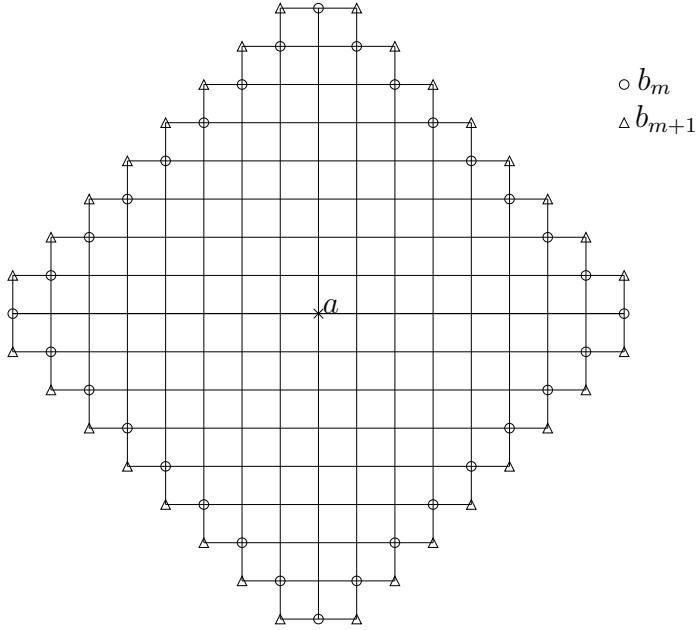}
 \caption{Finite lattice with boundary heights $(b_m,b_{m+1})$ fixed in a ground state configuration. \label{corner}} 
\end{figure}

Using Baxter's transfer matrix method \cite{refBaxter}, the Local State Probability (LSP) can be calculated exactly. The LSP $P(a| b,c)$ is the probability that a given site of the lattice has height $a$, with fixed boundary conditions in the ground state $(b,c)$. In order to calculate this quantity, we consider initially that the lattice is finite, with a shape as in Fig. [\ref{corner}], and fix the boundary heights to have the value they  would assume in a particular ground state configuration $(b,c)$ : $b_m= \left\{ \begin{array}{ccc} b & \textrm{if} &  m \textrm{ is odd} \\ c & \textrm{if} &  m \textrm{ is even} \end{array} \right.$ Finally, we take the limit $m\rightarrow \infty$ when the lattice becomes infinitely large, all boundary sites being infinitely far from the center site. 

Since this model is equivalent to a 4 fusion of the ABF model, the LSP has already been calculated in \cite{ref13}. In appendix \ref{LSP} we rederive it in this particular case. The exact expression for the LSP is given by the following expression :

\begin{eqnarray}
P(a|b,b+\eta) & = &  \frac{1}{L}\frac{ \theta_1(\pi \frac{a}{L},p)\theta_1(\pi/2,p^{L})}{\theta_1(\pi \frac{d}{L-2},p^{\frac{L}{L-2}})\theta_1(\pi \frac{3+\eta}{6},p^{\frac{L}{3}})}  \sum_{\Lambda-r-s=0 \textrm{ mod } 2} \cr & & \textrm{sin}\left(\frac{\pi (3+\eta)(\Lambda+1)}{6}\right)\textrm{sin}\left(\frac{\pi d r}{L-2}\right)\textrm{sin}\left(\frac{\pi a s}{L}\right)\chi_{\{\Lambda,r;s\}}(p^{L})
\label{LSP_expr}
\end{eqnarray}

where $c=b+\eta$,$d=(b+c)/2-1$ and $\chi_{\{\Lambda,r;s\}}$ are the branching coefficients of the coset $\frac{SU(2)_{k}\times SU(2)_{4}}{SU(2)_{k+4}} $ (cf appendix \ref{parafermions}) with $0  \leq \Lambda \leq   4, 1 \leq r \leq k + 1, 1 \leq s \leq  k + 5 $.

The appearance of the branching coefficients of the $\mathbb{Z}_{3}^{(2)}(k)$ parafermionic theory in the local state probability is quite interesting. The branching coefficients encodes the conformal data, and they should a priori be relevant for the model only at criticality. But here these objects apply to the model away from criticality. However surprising, this is a standard result for this general class of RSOS models in regime III \cite{ref15bis}. 

A lot of informations about the conformal field theory can be extracted from these branching coefficients. In particular the central charge and all primary conformal dimensions are accessible through the limit $q\rightarrow 0$ :

\begin{eqnarray*}
\chi_{\{\Lambda,r;s\}}(q) & = & \textrm{Tr}_{\{\Lambda,r;s\}}\left(q^{L_0-c/24}\right) \\
 & \sim_{q \rightarrow 0} & q^{\Delta_{r,s} + \delta -c/24}
\end{eqnarray*}

The expression (\ref{LSP_expr}) is valid for any value of the nome $p$. The probability that the central site of the lattice has height $a$ depends on the boundary conditions $(b,c)$, even in the thermodynamic limit. This means that there is long range order in this model away from criticality (for $0<p\leq1$). The regime III of this model is ordered.

Let us examine the expansion of the LSP around $p=0$. Since the branching coefficients behave as $\chi_{\{\Lambda,r;s\}}(t) \sim t^{-c/24+\Delta_{r,s}+\delta}(1+\mathcal{O}(t))$, the  LSP is dominated as $p \rightarrow 0$ by the branching coefficient of the identity operator ($\Delta = 0$). At $p = 0$ we find that the LSP becomes independant of the boundary condition $(b,b+\eta)$ :

\begin{equation}
 P(a|b,b+\eta) = P_a^{0} =  \frac{2}{L}\textrm{sin}\left(\frac{\pi a}{L}\right)^2 \nonumber
\end{equation}

Long range order disappears at $p=0$ : this is the signal of criticality. Moreover one can extract critical exponents by looking at the powers of $p$ in the LSP. The presence of branching coefficients in the LSP (\ref{LSP_expr}) forces critical exponents to be of the form :

\begin{equation}
 \beta = L \Delta \nonumber
\end{equation}

where $\Delta$ is a generic conformal dimension of the parafermionic theory $\mathbb{Z}_{3}^{(2)}(k)$.
$\alpha$ is taken as $2-\alpha=L$ in order to recover the standard relation between critical exponents and conformal dimensions :

\begin{eqnarray}
\Delta & = & \frac{\beta}{2-\alpha} 
\end{eqnarray}

This value of $\alpha$ determines immediatly the conformal dimension of the \emph{energy} operator through : 

\begin{equation}
\alpha = \frac{2- 4\Delta_{\epsilon}}{2- 2\Delta_{\epsilon}} \nonumber
\end{equation}

This leads to $\Delta_{\epsilon} = 1-\frac{2}{k+6}$, and it corresponds to the following neutral descendant of the doublet $\mathcal{D}_{1,3}$ :

\begin{eqnarray}
\Phi_{\epsilon} & = & \psi_{-\frac{2}{3}}^{\dagger}\mathcal{D}_{(1,3)}=\psi_{-\frac{2}{3}}\mathcal{D}_{(1,3)}^{\dagger} 
\end{eqnarray}

It is interesting to note that the pertubation of the second parafermionic theory $\mathbb{Z}_{3}^{(2)}(k)$ by this field $\psi_{-\frac{2}{3}}^{\dagger}\mathcal{D}_{(1,3)}$ has been studied in the continuouum limit \cite{reffusioncoset,refpert}. In particular it is known to be integrable in the continuum case : an infinite set of integrals of motion are preserved by this perturbation. The solvable model realizes this perturbation on the lattice.

\section{Discussions}

In this article solvable lattice models realizing at criticality the second parafermionic theory $\mathbb{Z}_{3}^{(2)}(k)$ are studied. A version with positive Boltzmann weights is presented, for the purpose of eventual numerical applications. The critical behavior is described by the parafermionic $\mathbb{Z}_{3}^{(2)}(k)$ conformal field theory, as can be seen through the evaluation of the Local State Probability. 

Away from criticality these models describe the second parafermionic theory $\mathbb{Z}_{3}^{(2)}(k)$ perturbed by the relevant field $\Phi_{\epsilon} = \Psi_{-2/3}^{\dagger}D_{1,3}$ with conformal dimension $\Delta_{\epsilon} =  1 - 2/(k+6)$. This perturbation by a neutral field should preserve the $\mathbb{Z}_3$ symmetry, but it remains hidden in both the Boltzmann weights and the ground states.

This is very different from the many solvable lattice models available for the first parafermions \cite{ref3}. For instance the N state spin model \cite{refparafermions_spin}, or the ABF model in regime II \cite{ref16}, in which the $\mathbb{Z}_N$ symmetry is explicit. The N state spin model has a $\mathbb{Z}_N$ invariant hamiltonian, while the ABF model in regime II realizes this symmetry at the level of ground states.

The model presented here for the $\mathbb{Z}_{3}^{(2)}(k)$ parafermions is not new, however for the general case $N \geq 5$ no lattice model is yet available. I hope to discuss this point in a future publication.

{\bf Acknowledgements:} Very useful discussions with Vl.~S.~Dotsenko are gratefully acknowledged.

\appendix

\section{Explicit expression for the Boltzmann weights}
\label{explicit_weights} 

Explicitly the weights are given by the following formulas. Some Boltzmann weights (\ref{minus1},\ref{minus2},\ref{minus3}) will be negative for $u=-1/4$, but it should be stressed that these signs are irrelevant and can be changed by a gauge transformation (cf section \ref{signs}).
The weights are parametrized in terms of the elliptic theta function : 
\begin{eqnarray}
 [u] & = & \theta_1(\frac{\pi u}{L})\\
\theta_1(u,p) & = & 2 p^{1/8} \textrm{sin}(u) \prod_{k=1}^{\infty} \left(1-2p^k \textrm{cos}2u + p^{2k}\right)\left(1-p^k \right)
\end{eqnarray}

which enjoys the following properties :
\begin{eqnarray}
[L-u] & = & [u] \\
 \phantom{L}[-u] & = & -[u]
 \end{eqnarray}

The following notations are used :

\begin{eqnarray}
 \mu & = & \pm 1 \\
a_{\mu} & = & \phantom{-}a \textrm{ if } \mu =1 \cr
& & -a \textrm{ if } \mu =-1 \\
\Delta_{\alpha}(x)& = & \left(\frac{[x+\alpha][x-\alpha]}{[x]^2}\right)^{1/2}
\end{eqnarray}

For all the 14 weights given here $\mu$ can take the values $\pm 1$. All remaining weights can be recovered from reflexion and rotational symmetry.

\begin{eqnarray}
W \left( \begin{array}{ll}
a & a + 2 \mu  \\
a + 2 \mu & a+ 4 \mu 
\end{array}  \right) & = & \frac{[1/2+u]}{[1/2]}\frac{[1+u]}{[1]}\frac{[3/2+u]}{[3/2]}\frac{[2+u]}{[2]} \\
W \left( \begin{array}{ll}
a & a +  \mu \phantom{2} \\
a +  \mu \phantom{2} & a+ 3 \mu 
\end{array}  \right) & = & \frac{[1/2+u]}{[1/2]}\frac{[1+u]}{[1]}\frac{[3/2+u]}{[3/2]}\frac{[a_{\mu}+3/2+u]}{[a_{\mu}+3/2]}\\
W \left( \begin{array}{ll}
a & a +  \mu\phantom{2}  \label{minus1}\\
a +  2\mu & a+ 3 \mu 
\end{array}  \right) & = & \frac{[1/2+u]}{[1/2]}\frac{[1+u]}{[1]}\frac{[3/2+u]}{[3/2]}\frac{[u]}{[2]} \Delta_{2}\left( a_{\mu}+3/2 \right)\\
W \left( \begin{array}{ll}
a & a +  2\mu  \\
a +  2\mu & a+ 2 \mu 
\end{array}  \right) & = & \frac{[1/2+u]}{[1/2]}\frac{[1+u]}{[1]}\frac{[a_{\mu} +1-u]}{[a_{\mu} +1]}\frac{[a_{\mu}+3/2-u]}{[a_{\mu}+3/2]}
\end{eqnarray}

\begin{eqnarray}
W \left( \begin{array}{ll}
a & a +  \mu \phantom{2} \\
a +  \mu \phantom{2} & a+ 2 \mu 
\end{array}  \right) & = & \frac{[1/2+u]}{[1/2]}\frac{[1+u]}{[1]} \left( \frac{[1-u]}{[1]}\frac{[3/2+u]}{[3/2]} \right. \cr
& + & \left.  \frac{[u]}{[2]}\frac{[1/2+u]}{[1/2]} \frac{[a_{\mu}-1/2][a_{\mu}+5/2]}{[a_{\mu}+1/2][a_{\mu}+3/2]}  \right) \\
W \left( \begin{array}{ll}
a & a \phantom{+ 2\mu}  \\
a +  2\mu & a+ 2 \mu 
\end{array}  \right) & = & -\frac{[1/2+u]}{[1/2]}\frac{[1+u]}{[1]}\frac{[1/2-u]}{[3/2]}\frac{[u]}{[2]}  \Delta_{1}(a_{\mu}+1/2)  \cr
 & &  \Delta_{1}(a_{\mu}+3/2)\Delta_{2}\left( a_{\mu}+1 \right)  \\
W \left( \begin{array}{ll}
a & a   \\
a + \mu\phantom{2}  & a+ 2 \mu 
\end{array}  \right) & = &  \frac{[1/2+u]}{[1/2]}\frac{[1+u]}{[1]}\frac{[u]}{[1/2]}\frac{[a_{\mu}+1+u]}{[a_{\mu}+1]} \cr & & \left(\frac{[1/2][1]}{[3/2][2]}  \frac{[a_{\mu}+5/2][a_{\mu}-1]}{[a_{\mu}+3/2][a_{\mu}+1]} \right)^{1/2}  \label{minus2}\\
W \left( \begin{array}{ll}
a & a +2 \mu  \\
a +  2\mu & a+  \mu 
\end{array}  \right) & = & \frac{[1/2+u]}{[1/2]} \frac{[a_{\mu}+1-u]}{[a_{\mu}+1]} \frac{[a_{\mu}+1/2-u]}{[a_{\mu}+1/2]} \cr  & & \frac{[a_{\mu}+3/2-u]}{[a_{\mu}+3/2]} 
\end{eqnarray}

\begin{eqnarray}
W \left( \begin{array}{ll}
a  \phantom{+ 2\mu}& a - \mu \phantom{2}  \\
a   \phantom{+ 2\mu } & a+  \mu \phantom{2} 
\end{array}  \right) & = & -\frac{[1/2+u]}{[1/2]}\frac{[u]}{[2]}\frac{[a_{\mu}+u]}{[a_{\mu}-1/2]}\frac{[a_{\mu}+1/2-u]}{[a_{\mu}+1/2]} \cr  & & \left(\frac{[3/2][2]}{[1][1/2]} \right)^{1/2} \left(\frac{[a_{\mu}+3/2][a_{\mu}-3/2]}{[a_{\mu}+1][a_{\mu}]}\right)^{1/2}  \\
W \left( \begin{array}{ll}
a & a  \phantom{+ 2\mu} \\
a \phantom{+ 2\mu}  & a+ \mu  \phantom{2}
\end{array}  \right) & = &\frac{[1/2+u]}{[1/2]}\frac{[a_{\mu}+1/2+u]}{[a_{\mu}+1/2]}\left( \frac{[1/2-u]}{[1/2]}\frac{[1+u]}{[1]} \right. \cr
& +& \left.  \frac{[u][1/2+u]}{[1/2]^2}\frac{[3/2]}{[2]}\frac{[a_{\mu}-1][a_{\mu}+3/2]}{[a_{\mu}+1][a_{\mu}-1/2]} \right)\\
W \left( \begin{array}{ll}
a  \phantom{+2\mu}& a \phantom{+2\mu} \\
a+\mu\phantom{2}   & a+  \mu\phantom{2} 
\end{array}  \right) &= &\frac{[u][1/2+u]}{[1/2]^2}\left( \frac{[a_{\mu}-1][a_{\mu}][a_{\mu}+1][a_{\mu}+2]}{[a_{\mu}-1/2][a_{\mu}+1/2]^2[a_{\mu}+3/2]}\right)^{1/2} \cr 
\left(  \frac{[1/2-u]}{[3/2]}\frac{[1+u]}{[1/2]}  \right.& + & \left. \frac{[u]}{[1]}\frac{[1/2+u]}{[1/2]}\frac{[3/2]}{[2]}\frac{[a_{\mu}-1/2][a_{\mu}+3/2]}{[a_{\mu}][a_{\mu}+1]} \right) \label{minus3}\\
W \left( \begin{array}{ll}
a & a +2 \mu   \\
a +2 \mu  & a 
\end{array}  \right) & = &  \frac{[a_{\mu}-u]}{[a_{\mu}]}\frac{[a_{\mu}+1/2-u]}{[a_{\mu}+1/2]}\frac{[a_{\mu}+1-u]}{[a_{\mu}+1]}\cr & & \frac{[a_{\mu}+3/2-u]}{[a_{\mu}+3/2]}
\end{eqnarray}

\begin{eqnarray}
W \left( \begin{array}{ll}
a & a + \mu   \\
a + \mu  & a 
\end{array}  \right) & = &  \frac{[a_{\mu}-u]}{[a_{\mu}]}\frac{[a_{\mu}+1/2-u]}{[a_{\mu}+1/2]}\left\{\frac{[1+u]}{[1]}\frac{[1/2-u]}{[1/2]} + \left( \frac{[3/2]^2[1]}{[1/2]^2[2]} \right. \right. \cr
&  -& \left. \left.  \frac{[2][a_{\mu}+1/2]^2}{[1][a_{\mu}-1/2][a_{\mu}+3/2]}\right)  \frac{[u]}{[1]}\frac{[1/2+u]}{[1/2]} \right\} \\
W \left( \begin{array}{ll}
a & a \phantom{+ \mu}    \\
a \phantom{+ \mu} & a 
\end{array}  \right) & = &  \frac{[a_{\mu}-u]}{[a_{\mu}]}\frac{[a_{\mu}+1/2-u]}{[a_{\mu}+1/2]}\frac{[a_{\mu}+1+u]}{[a_{\mu}+1]}\frac{[a_{\mu}+1/2+u]}{[a_{\mu}+1/2]} \cr
 & + & \frac{[u]}{[1]}\frac{[1/2+u]}{[1/2]}\frac{[a_{\mu}-3/2][a_{\mu}+1]}{[a_{\mu}-1/2]^2} \cr & & \left\{ \frac{[a_{\mu}-u][a_{\mu}+1/2+u]}{[a_{\mu}+1/2]^2}\frac{[1]^3}{[1/2]^2[2]} \right. \cr
 & - & \left. \frac{[1+u][1/2-u]}{[2][3/2]} \frac{[a_{\mu}+1/2][a_{\mu}-1][a_{\mu}-2]}{[a_{\mu}+1]^2[a_{\mu}]} \right\}
\end{eqnarray}

\section{The second parafermionic theory $\mathbb{Z}_3^{(2)}(k)$}
\label{parafermions} 

The details of the second parafermionic theories $\mathbb{Z}_3^{(2)}(k)$ can be found in \cite{ref3}. These CFT possess an infinite symmetry generated by parafermionic currents. The Virasoro generators form a subalgebra of this enhanced algebra. The second parafermionic theories are realized by the following coset construction :

\begin{equation}
\mathbb{Z}_3^{(2)}(k) = \frac{SU(2)_{k}\times SU(2)_{4}}{SU(2)_{k+4}} \nonumber
\end{equation}

and possess a central charge :

\begin{equation}
 c=2\left(1-\frac{12}{(k+2)(k+6)} \right) \nonumber
\end{equation}

The chiral algebra is made of $2$ parafermionic currents $\Psi,\Psi^{\dagger}$ of dimension $4/3$, and $\mathbb{Z}_3$ charge $\pm 1$. They obey the following operator product expansion : 

\begin{eqnarray}
\Psi \times \Psi^{\phantom{\dagger}} & \rightarrow & \Psi^{\dagger}\\
\Psi \times \Psi^{\dagger} & \rightarrow & \mathbb{I}
\end{eqnarray}

The parafermionic algebra primaries of the second $\mathbb{Z}_{3}$ conformal theory are labeled by 2 integers ($r,s$), with conformal dimension :

\begin{eqnarray}
\Delta_{r,s} & = & \frac{ \left( r (k+6) - s (k+2)  \right)^2- 16 }{16(k+2)(k+6)} + \frac{m(4-m)}{48} \\
m & = & s-r \quad \textrm{mod } 4 \\
1 \leq & r & \leq k+1 \cr
1 \leq & s & \leq k+5  \nonumber
\end{eqnarray}

The value of $m$ labels different sectors of the theory. \emph{Singlets} ($m=0$) are neutral fields, while \emph{doublets} ($m=2$) have a $\mathbb{Z}_3$ charge $\pm1$. Finally $m=1,3$ correspond to \emph{disorder} fields.

\begin{figure}
 \centering
 \psfrag{S}[0][0][1][0]{$\mathcal{S}$}
 \psfrag{Psi_{-1/3}S}[0][0][1][0]{$\Psi_{-1/3}\mathcal{S}$}
 \psfrag{Psi_{-1/3}^{dagger}S}[0][0][1][0]{$\Psi_{-1/3}^{\dagger}\mathcal{S}$}
 \psfrag{q=0}[0][0][1][0]{$q=0$}
 \psfrag{q=+1}[0][0][1][0]{$q=+1$}
 \psfrag{q=-1}[0][0][1][0]{$q=-1$}
 \psfrag{1/3}[0][0][1][0]{$1/3$}
 \includegraphics[scale=0.7]{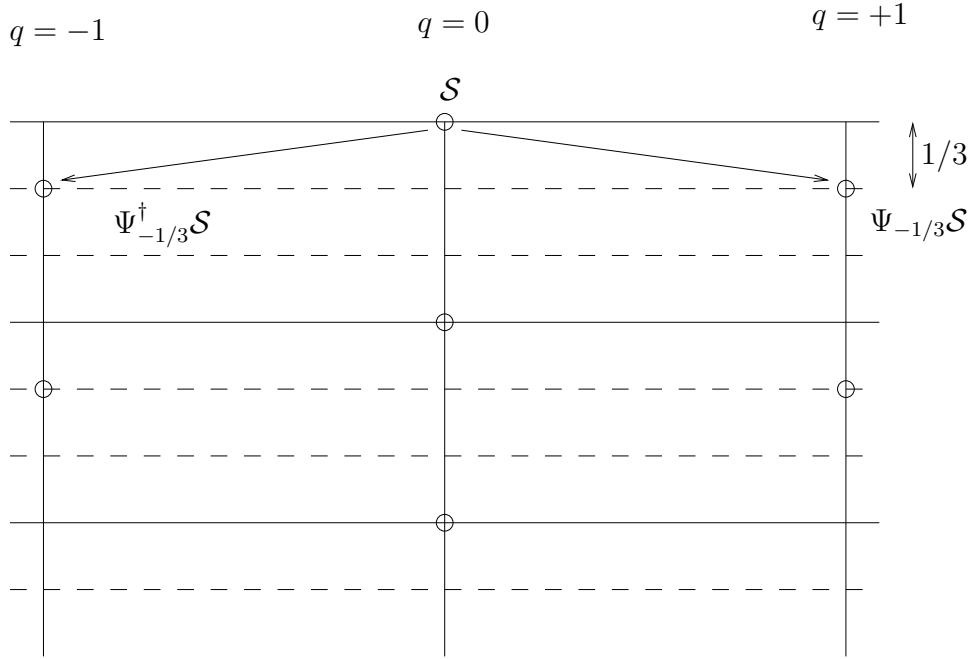}
 \caption{Structure of a singlet module. Arrows depict the first (charged) descendants of the primary singlet $\mathcal{S}$, obtained by action of the parfermionic mode operators. \label{singlet}}
\end{figure}

\begin{figure}
 \centering
 \psfrag{D}[0][0][1][0]{$\mathcal{D}$}
 \psfrag{D^{dagger}}[0][0][1][0]{$\mathcal{D}^{\dagger}$}
 \psfrag{Psi_{-2/3}D^{dagger}}[0][0][1][0]{$\Psi_{-2/3}\mathcal{D}^{\dagger},\Psi_{-2/3}^{\dagger}\mathcal{D}$}
 \psfrag{q=0}[0][0][1][0]{$q=0$}
 \psfrag{q=+1}[0][0][1][0]{$q=+1$}
 \psfrag{q=-1}[0][0][1][0]{$q=-1$}
 \psfrag{2/3}[0][0][1][0]{$2/3$}
 \includegraphics[scale=0.7]{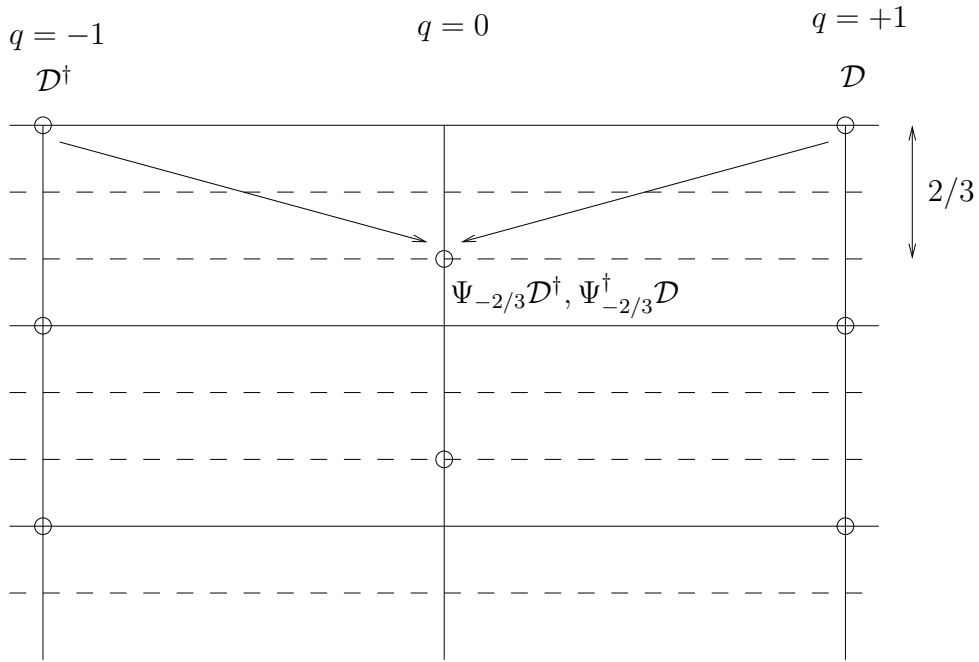}
 \caption{Structure of a doublet module. Arrows depict the first (neutral) descendants of the primary doublet $\mathcal{D},\mathcal{D}^{\dagger}$, obtained by action of the parfermionic mode operators. \label{doublet}}
\end{figure}

In each sector $(r,s)$, descendants will have a conformal dimension $\Delta_{r,s} + \delta + \mathbb{N}$  depending on their $\mathbb{Z}_3$ charge (cf Fig. \ref{singlet},\ref{doublet}):
\begin{eqnarray*}
 \delta = 0 \textrm{ or } 1/3 & & \textrm{in a singlet module}\\
 \delta = 0 \textrm{ or } 2/3 & & \textrm{in a doublet module}\\
 \delta = 0 \textrm{ or } 1/2 & & \textrm{in a disorder module}\\
\end{eqnarray*}

In particular the conformal dimension of the neutral field $\epsilon = \psi_{-\frac{2}{3}}^{\dagger}\mathcal{D}_{(1,3)}=\psi_{-\frac{2}{3}}\mathcal{D}_{(1,3)}^{\dagger}$ is :

\begin{eqnarray}
\Delta_{\epsilon} & = & \frac{k+4}{k+6} = 1 -\frac{1}{L}
\end{eqnarray}

The branching coefficients of the corresponding coset $\frac{SU(2)_{k}\times SU(2)_{4}}{SU(2)_{k+4}} $ appear naturally in the LSP. They are expressed in terms of level 4 string functions of $SU(2)$  $c_m^{\Lambda}$ \cite{refDiFr} :

\begin{equation}
\chi_{\{\Lambda,r;s\}} = \textrm{Tr}_{\{\Lambda,r;s\}}\left(q^{L_0-c/24}\right)=\sum_{m=0}^{7} c_m^{\Lambda} F_m(q)
\end{equation}

\begin{equation}
F_m(q)= \sum_{n=-\infty}^{\infty} \delta_{m,m'} q^{\beta_{r-2n(k+6),s}} - \delta_{m,m''} q^{\beta_{-r-2n(k+6),s}}
\end{equation}

and
\begin{equation}
\beta_{r,s} = \frac{((k+6)r - (k+2)s)^2}{16(k+2)(k+6)}
\end{equation}

\begin{eqnarray*}
m'=r-2n(k+6) - s & \textrm{mod} & 8 \\
m''= -r - 2n(k+6) - s& \textrm{mod} & 8
\end{eqnarray*}

In particular as $q\rightarrow 0$ they behave as :

\begin{equation}
\chi_{\{\Lambda,r;s\}}(q) \sim q^{\Delta_{r,s}+\delta_{\Lambda,r,s}-c/24}(1+\mathcal{O}(t))
\end{equation}

where $\delta_{\Lambda,r,s}$ depends on the subsector of the primary field $\Phi_{r,s}$ : 

\begin{itemize}
 \item if $r-s=0$ mod $4$ : then $m'=0,4$ mod $8$. Using $c_m^l=c_{4-m}^{4-l}$ we can always take $m'=0$

Then we get :
\begin{eqnarray*}
q^{2/24}c_0^0 & \sim & 1 \\
q^{2/24}c_0^2 & \sim & q^{1/3} \\
q^{2/24}c_0^4 & \sim & q^{1} 
\end{eqnarray*}

reproducing the structure of the module of a singlet : neutral descendants of a singlet have a conformal dimension in  $\Delta_{r,s} + \mathbb{N}$, while charged descendants have a dimension in $\Delta_{r,s} + 1/3 + \mathbb{N}$

\item if $r-s=2$ mod $4$ : then $m'=2,6$ mod $8$. Using $c_m^l=c_{-m}^{l}$ we can always take $m'=2$

In this case we get :
\begin{eqnarray*}
q^{2/24}c_2^2 & \sim & q^{1/12} \\
q^{2/24}c_2^0 & \sim & q^{1/12+2/3} 
\end{eqnarray*}

which is precisely the structure of the module of a doublet : charged descendants correspond to $\delta = 0$, and neutral descendants to $\delta = 2/3$.

\item if $r-s=\pm 1$ mod $4$ : then $m'=1,3,5,7$ mod $8$. We can always take $m'=1$

\begin{eqnarray*}
q^{2/24}c_1^1 & \sim & q^{1/16} \\
q^{2/24}c_1^3 & \sim & q^{1/16+1/2} 
\end{eqnarray*}

describing the disorder sector.

\end{itemize}

\section{Local state probability}
\label{LSP}

Since this model is strictly equivalent to a 4 fusion of the ABF model, the LSP has already been calculated in \cite{ref13}. In this appendix this result is rederived in a slightly different manner. In particular a simpler form is given for the non restricted one dimensional sum, more in the spirit of \cite{ref15bis}.

Using the corner transfer matrix method \cite{refBaxter}, one obtains for the LSP : 
\begin{equation}
P(a| b,c) = \lim_{m\rightarrow \infty} P_m(a,b_m,b_{m+1})
\end{equation}

with
\begin{eqnarray}
 P_m(a,b,c) & = & S_m^{-1} u_a X_m(a,b,c;x^{-2\lambda})  \cr
u_a & = &   E(x^a,x^L)  \cr
S_m & = & \sum_a u_a X_m(a,b,c) \cr
X_m(a,b,c;q) & = & \sum_{a_2,...,a_m}q^{\sum_{j=1}^{m} jH(a_j,a_{j+1}.a_{j+2})}
\end{eqnarray}

where the sum is taken on all admissible paths $(a_2,...a_m)$ from $a_1=a$ to $a_{m+1}=b_{m+1}$. The problem is now to evaluate the one-dimensional configuration sum $X_m$, and then the normalization $S_m$.

The restricted 1D sum $X_m(a |b,c;q)$ is uniquely determined by the recurrence :

\begin{eqnarray}
X_m(a,b,c;q) & = & \sum_{a_m}'' X_{m-1}(a,a_m,b)q^{mH(a_m,b,c)} \\
X_0(a,b,c;q) & = & \delta_{a,b}
\end{eqnarray}

where the sum $\sum''$ is taken over $a_m$ such that the pair $(a_m,b)$ is admissible, and $H(a,b,c)=H(b-a,c-a)$ has the following form : 

\begin{equation}\begin{array}{|c||c|c|c|c|c|}
\hline
 H_{\mu,\nu} & -2   & -1  & 0   & +1  & +2\\ \hline \hline
     -2      &  2   & 3/2 & 1   & 1/2 & 0   \\ \hline
     -1      &  3/2 & 1   & 1/2 & 0   & 1/2 \\ \hline
      0      &  1   & 1/2 & 0   & 1/2 & 1   \\ \hline
     +1      &  1/2 & 0   & 1/2 & 1   & 3/2 \\ \hline
     +2      &  0   & 1/2 & 1   & 3/2 & 2   \\ \hline
\end{array}
\end{equation}

Let's consider initially the non restricted one-dimensional configuration sum : 
\begin{eqnarray}
f_m(a,b,c;q) & = & \sum_{a_m}{}' f_{m-1}(a,a_m,b)q^{mH(a_m,b,c)} \\
f_0(a,b,c;q) & = & \delta_{a,b}
\end{eqnarray}

where the sum $\sum'$ is now taken over $a_m$ such that the pair $|a_m-b|=\pm 2,\pm1,0$ (i.e. we forget the bounds over the heights). This quantity is invariant under translation, so that it depends only on $(\gamma,\eta)=(b-a,c-b)$

The solution can be expressed in terms of q-multinomial coefficients $\left[ \begin{array}{c} m \\ k_1...k_n \end{array}\right]$: 

\begin{equation}\left[ \begin{array}{c} m \\ k_1...k_n \end{array}\right] = \frac{(q)_m}{(q)_{k_1}...(q)_{k_n}} \quad  \quad \left(\sum_i k_i = m, \quad k_i \geq 0 \right) \end{equation} 

with \begin{equation}(q)_n= \prod_{i=1}^{n}(1-q^i) \quad  \quad \left(  n \geq 0 \right) \end{equation}

The unrestricted 1D sum is then of the form :

\begin{equation}f_m(\gamma,\eta;q) = \sum_{k}{}^* q^{\mathcal{Q}(k)+\sum_i k_i H(i,\eta)} \left[ \begin{array}{c} m \\ k_i \end{array}\right] \end{equation}

where $k=(k_{-2},k_{-1},k_0,k_1,k_2)$, the sum $\sum_{k}^*$ is taken over $k$ such that :

\begin{eqnarray*}
k_i & \geq & 0 \\
\sum_{i=-2}^2 k_i & = & m\\  
\sum_{i=-2}^2 i k_i & = & \gamma 
\end{eqnarray*}

and $\mathcal{Q}(k)$ is the following quantity :

\begin{equation}\mathcal{Q}(k) = \sum_i k_i(k_i-1) - 2k_2k_{-2}+(k_2-k_{-2})(k_1-k_{-1}) + k_0(k_1+k_{-1}) \end{equation}

This can be simplified using the constraints on $k$:

\begin{equation}\mathcal{Q}(k) = \gamma^2/4 - m + \left( k_0+\frac{k_1+k_{-1}}{2}\right)^2  + \frac{k_1^2+k_{-1}^2}{2} \end{equation}

The restricted 1D sum is then the usual sum over all reflexions along the boundaries. We get :
\begin{eqnarray}
X_m(a,b,c;q) & = & Y_m(a,b,c;q) - q^{-a} Y_m(-a,b,c;q)\\
Y_m(a,b,c;q) & = & \sum_{k=-\infty}^{\infty} q^{-2(k^2 L+k a)+ k L}  f_m(b-a - 2 k L,b-c;q)
\end{eqnarray}

It is straightforward to check that this formula for $X_m$ obey :
\begin{eqnarray*}
X_m(a,0,2;q) & = & 0 \\
X_m(a,L,L-2;q) & = & 0 
\end{eqnarray*}

The other cases are more involved. For $\eta \in \{\pm2, \pm1,0 \}$ the following equations are satisfied by $X_m$ : 

\begin{eqnarray*}
q^{mH(1,\eta)}X_m(a,1/2,3/2;q) + q^{mH(2,\eta)}X_m(a,-1/2,3/2;q) & = & 0   \\
q^{mH(0,\eta)}X_m(a,1,1;q) + q^{mH(1,\eta)}X_m(a,0,1;q) + q^{mH(2,\eta)}X_m(a,-1,1;q)  & = & 0   \\
q^{mH(-1,\eta)}X_m(a,3/2,1/2;q) + q^{mH(0,\eta)}X_m(a,1/2,1/2;q) & + & \\ q^{mH(1,\eta)}X_m(a,-1/2,1/2;q) +q^{mH(2,\eta)}X_m(a,-3/2,1/2;q) & = & 0  
\end{eqnarray*}

All this relations ensure that the restricted recurrences are satisfied by $X_m$.

The limit $m \rightarrow \infty$ of of $f_m(\gamma,\eta;q)$ has the following form can be taken by expressing $k_2,k_{-2}$ in terms of the other variables $k_1,k_0,k_{-1}$ : 

\begin{eqnarray*}
4k_2 & = & \gamma + 2m - (3k_1+2k_0+k_{-1}) \\
4k_{-2} & = & - \gamma + 2m - (k_1+2k_0+3 k_{-1})
\end{eqnarray*}
 
The sum is now on $(k_1,k_0,k_{-1})$ such that $k_2$ and $k_{-2}$ are integers :  $2m +2 k_0 = k_1 -k_{-1} + \gamma \quad \textrm{mod } 4$

Moreover the $\eta$ dependent term  $ \sum_i k_i H(i,\eta)$ takes the form :

\begin{eqnarray*}
m+\gamma/2 & \textrm{if} & \eta=2 \\
m+\gamma/4 - \frac{3k_1 +2k_0 +k_{-1}}{2} & \textrm{if} & \eta=1 \\
m - \frac{k_1 +2k_0 +k_{-1}}{2} & \textrm{if} & \eta=0 \\
m-\gamma/4 - \frac{k_1 +2k_0 +3k_{-1}}{2} & \textrm{if} & \eta=-1 \\
m-\gamma/2  & \textrm{if} & \eta=-2 
\end{eqnarray*}

\begin{eqnarray}
 \lim_{m \rightarrow \infty} f_m(\gamma,\eta) & = & q^{\gamma(\gamma +\eta)/4}\frac{1}{(q)_{\infty}}  \sum_{k_1,k_0,k_{-1}}{}^* q^{Q_\eta(k)}\frac{1}{(q)_{k_1}(q)_{k_0}(q)_{k_{-1}}} \\ 
 & = & q^{\gamma(\gamma +\eta)/4} \frac{1}{(q)_{\infty}} \Phi_{(s_0,\eta)}(q) \nonumber
\end{eqnarray}

where $Q_{\eta}(k)$ is given by :
\begin{equation} \left( k_0+\frac{k_1+k_{-1}}{2}\right)^2  + \frac{k_1^2+k_{-1}^2}{2} - 
\left\{\begin{array}{ccc}
0 & \textrm{if} & \eta=2 \\
  \frac{3k_1 +2k_0 +k_{-1}}{2} & \textrm{if} & \eta=1 \\
 \frac{k_1 +2k_0 +k_{-1}}{2} & \textrm{if} & \eta=0 \\
 \frac{k_1 +2k_0 +3k_{-1}}{2} & \textrm{if} & \eta=-1 \\
0  & \textrm{if} & \eta=-2 
\end{array}\right.\end{equation}

$\sum {}^*$ stands for the restriction $2 k_0 + k_1 - k_{-1} = s_0 \, \textrm{mod} \, 4 $, and $s_0 = 2m - \gamma \, \textrm{mod} \, 4$.

Up to a multiplicative factor, the functions $\Phi_{(s_0,\eta)}(q)$ are precisely the branching functions of the $c=1$ $\mathbb{Z}_4$ parafermions \cite{refMcCoy}: 

\begin{equation} \Phi_{(s_0,\eta)}(q) = q^{1/24 -(2+\eta)(2-\eta)/48 }b_{2s_0-2-\eta}^{2+\eta} \end{equation}

The branching functions $b_m^l$ are defined for $l$ $\textrm{mod} \, 4$, and $m$  $\textrm{mod} \, 8$, $l-m$ even, and enjoy the following symmetries :

\begin{eqnarray*}
 b_m^l & = & b_{-m}^l \\
 b_m^l & = & b_{m+8}^l \\
 b_m^l & = & b_{4-m}^{4-l} 
\end{eqnarray*}

Finally we get for the limit of $f_m$ :

\begin{eqnarray} \lim_{m \rightarrow \infty} f_m(\gamma,\eta) &  = & q^{\gamma(\gamma +\eta)/4}\frac{1}{(q)_{\infty}} q^{1/24 -(2+\eta)(2-\eta)/48 }b_{2s_0-2-\eta}^{2+\eta} \cr
 & = & q^{\gamma(\gamma +\eta)/4} \eta(q)^{-1} q^{2/24 - (4-\eta^2)/48} b_{2s_0-2-\eta}^{2+\eta} \cr
 & = & q^{\gamma(\gamma +\eta)/4}  q^{\eta^2/48} c_{2s_0-2-\eta}^{2+\eta}
 \end{eqnarray}

where $c_m^l = \eta(q)^{-1} b_m^l$ is a level 4 string function of $SU(2)$. We now impose the ground state $(b,\eta)$ at the boundary :

\begin{equation}
b_m = \left\{\begin{array}{ccc}
b & \textrm{if} & m \textrm{ is odd} \\
c = b+\eta & \textrm{if} & m \textrm{ is even} 
\end{array}\right.
\end{equation}

and compute the thermodynamic limit $m \rightarrow \infty$ of $f_m(b_m-a,b_{m+1}-b_m;q)$ (in the case $m$ odd for instance, but the result does not depend on the way the limit is taken ) :

Let's say that in the case $m$ odd we have a certain $s_0= 2 -(b-a)$ $\textrm{mod} \, 4$. Then for $m$ even we will have $s_0' = 0 - (b + \eta -a) = s_0 - (2+\eta)$. 
\begin{eqnarray*}
\lim_{m\rightarrow \infty}f_m(b_m-a,b_{m+1}-b_m;q) = q^{(b-a)(b+\eta-a)/4}q^{\eta^2/48}c_{2s_0-2-\eta \phantom{+4}}^{2+\eta} & \textrm{if} & m \textrm{ is odd} \\
\lim_{m\rightarrow \infty}f_m(b_m-a,b_{m+1}-b_m;q) = q^{(b + \eta -a)(b-a)/4}q^{\eta^2/48}c_{2s_0-2-\eta + 4}^{2-\eta} & \textrm{if} & m \textrm{ is even} 
 \end{eqnarray*}

But the symmetries of the branching coefficients ($b_m^l = b_{m-4}^{4-l} $) ensure that $c_{2s_0-2-\eta + 4}^{2-\eta} = c_{2s_0-2-\eta }^{2+\eta}$

\begin{equation}
\lim_{m\rightarrow \infty}f_m(b_m-a,b_{m+1}-b_m;q)   =  q^{(d-a+1)^2/4}q^{-\eta^2/24}c_{m'}^{2+\eta} 
\end{equation}

where $m'=2a-2d\, \textrm{mod}\, 8$, $d=(b+c)/2-1 = b + \eta/2 -1$

Recalling that $L=k/2+3$, we recognize in $\lim_{m\rightarrow \infty} X_m(a,b_m,b_{m+1})$ a $\frac{SU(2)_{k}\times SU(2)_{4}}{SU(2)_{k+4}} $ branching function (cf appendix \ref{parafermions}) :

\begin{eqnarray}
\lim_{m\rightarrow \infty} X_m(a,b_m,b_{m+1}) & = & q^{(b-a)(c-a)/4 + \eta^2/48- \frac{(a(L-2)-dL)^2}{4L(L-2)}} \chi_{\{2+\eta, 2 d;2 a\}} \cr
  & = & q^{(d+1-a)^2/4 - \eta^2/24- \frac{(a(L-2)-dL)^2}{4L(L-2)}} \chi_{\{2+\eta, 2 d;2 a\}}
\end{eqnarray}

As usual in regime III the principal specialization of the character identity arising from the coset construction comes very handy to normalize the LSP. 
 
\begin{eqnarray}
P(a|b,\eta) & = & \frac{\chi_{\{\Lambda,2d;2a\}} \chi_{2a-1}^{(k+4)}}{\chi_{\Lambda}^{(4)} \chi_{2d-1}^{(k)}} \\
 & = &  x^{(d+1-a)^2/4 - \eta^2/24- \frac{(a(L-2)-dL)^2}{4L(L-2)}} \frac{Q(x^{1/2})^2}{Q(x)}\\
 &   & \frac{E(x^a,x^L)}{E(x^d,x^{L-2})E(x^{(3+\eta)/2},x^3)} \chi_{\{2+\eta,  2d; 2a\}}(x) 
\end{eqnarray}

where we introduced :
\begin{eqnarray*}
 E(z,x)& = & \prod_{n=1}^{\infty}\left(1-zx^{n-1}\right)\left(1-z^{-1}x^{n}\right)\left(1-x^{n}\right) \\
 & = & \sum_{k=-\infty}^{\infty} x^{k(k-1)/2} (-z)^k \\
Q(x) & = & \prod_{n=1}^{\infty}\left(1-x^n \right)
\end{eqnarray*}

In order to study the critical behavior of the model, described by the LSP around $p=0$, it is useful to go back to the variable $p$ rather than $x$. Then again, having identified in the LSP a ratio of characters is of great help since modular transformations are readily available. Finally we get the following expression :

\begin{eqnarray}
P(a|b,c) & = &  \frac{1}{L}\frac{ \theta_1(\pi \frac{a}{L},p)\theta_1(\pi/2,p^{L})}{\theta_1(\pi \frac{d}{L-2},p^{\frac{L}{L-2}})\theta_1(\pi \frac{3+\eta}{6},p^{\frac{L}{3}})}  \sum_{\Lambda-r-s=0 \textrm{ mod } 2} \cr & & \textrm{sin}\left(\frac{\pi (3+\eta)(\Lambda+1)}{6}\right)\textrm{sin}\left(\frac{\pi d r}{L-2}\right)\textrm{sin}\left(\frac{\pi a s}{L}\right)\chi_{\{\Lambda,r;s\}}(p^{L})
\end{eqnarray}

with : 
\begin{eqnarray*}
d & = & \frac{b+c}{2}-1 \\
\eta & = & b-c
\end{eqnarray*}

and the sum in the r.h.s. is over :

\begin{eqnarray*}
 0 & \leq \Lambda \leq &  4 \\
 1 & \leq r \leq &  2L-5 = k + 1 \\
 1 & \leq s \leq &  2L-1 = k + 5 
 \end{eqnarray*}

\end{document}